\documentclass[letter,traditabstract]{aa} 
\usepackage{graphicx}
\usepackage{color}
\usepackage{microtype}
\usepackage{natbib}
\bibpunct{(}{)}{;}{a}{}{,} 
\usepackage{txfonts}

\def \approxgt{\mathrel{\hbox{\rlap{\lower.55ex \hbox {$\sim$}}
        \kern-.3em \raise.4ex \hbox{$>$}}}}
\def \approxlt{\mathrel{\hbox{\rlap{\lower.55ex \hbox {$\sim$}}
        \kern-.3em \raise.4ex \hbox{$<$}}}}
\def\xte{{\em RXTE}}

\def\gx{GX\,3+1}

\def\xmm{{\em XMM-Newton}}

\def\suzaku{{\em Suzaku}}
\def \xmm {{\em XMM--Newton}}

\def \sax {{\em BeppoSAX}}

\def \sax{{\it BeppoSAX}}
\def \countsec{\hbox{counts s$^{-1}$}}

\begin{document}
\titlerunning{A relativistic iron line from \gx\ }

\title{A relativistic iron emission line from the neutron star low-mass X-ray binary \gx\ }

   \author{S. Piraino
          \inst{1,2}
      \and A. Santangelo
           \inst{2}
            \and P. Kaaret
            \inst{3}
      \and B. M\"uck
       \inst{2}
  \and A. D'A\`i
        \inst{4}
      \and T. Di Salvo
            \inst{4}     
           \and R. Iaria
            \inst{4}
            N. Robba
            \inst{4}
          \and L. Burderi
            \inst{5}
          \and E. Egron
            \inst{5}            
          }

   \institute{INAF-IASF di Palermo, via Ugo La Malfa 153, 90146 Palermo, Italy; \email{santina.piraino@ifc.inaf.it}
              \and 
             Institut f\"ur Astronomie und Astrophysik, Kepler Center for Astro and Particle Physics, Sand 1, 72076 T\"ubingen, Germany
              \and  
              Department of Physics and Astronomy University of Iowa, Iowa City, IA 52242 USA
              \and
              Dipartimento di Fisica, Universit\`a degli Studi di Palermo, via Archirafi, 90100 Palermo, Italy
            \and
            Dipartimento di Fisica, Universit\`a degli Studi di Cagliari, SP Monserrato-Sestu, KM 0.7, Monserrato, Italy
             }

   \date{Submitted: 26 March 2012  - Accepted: 8 May 2012 }

\abstract{We present the results of a spectroscopic study of the Fe K$\alpha$ emission of the persistent neutron-star atoll low-mass X-ray binary and type I X--ray burster \gx with the EPIC-PN on board \xmm. The source shows a flux modulation over several years and we observed it during its fainter phase, which corresponds to an X-ray luminosity of $L_X\sim 10^{37}$ ergs s$^{-1}$. When fitted with a two-component model, the X-ray spectrum shows broad residuals at $\sim 6-7$ keV that can be ascribed to an iron $K_{\alpha}$ fluorescence line. In addition, lower energy features are observed at $\sim 3.3$ keV, $\sim 3.9$ keV and might originate from Ar~XVIII and Ca XIX. The broad iron line feature is well fitted with a relativistically smeared profile. This result is robust against possible systematics caused by instrumental pile-up effects.  Assuming that the line is produced by reflection from the inner accretion disk, we infer an inner disk radius of $\sim 25 R_g$ and a disk inclination of $35^\circ<i<44^\circ$.}


\maketitle

\section{Introduction}
 
\label{sec:intro}

Spectra of accreting compact objects, such as active galactic nuclei  and X-ray binary systems containing  stellar mass black holes or  weakly magnetized neutron stars (NS), often exhibit \emph{broad} emission lines (FWHM up to $\sim1$~keV) at $E\sim6.4-6.97$~keV, identified with K$_{\alpha}$ radiative transitions of iron at different ionization states. These broad lines are thought to originate from reprocessed emission that is relativistically smeared, in the inner parts of the accretion disk illuminated by the primary Comptonized spectrum. This process  produces different spectral signatures, including a reflection hump that peaks between 20-30 keV, emission lines, and absorption edges (see  \citealt{fabian00} for a review). Owing to their relatively high abundance and fluorescence yield, the iron lines are the best suited features to diagnose the accretion flows close to the compact object, because the predicted line profile depends on the geometry - inclination and inner radius - of the system \citep{fabian89}. For NS systems, the inner disk radius sets an upper limit to the radius of the star, which can constrain the NS equation of state. 

\citet{piraino00} first suggested a relativistic origin of the broad iron line observed in the NS system 4U 1728-34. 
Recently, thanks to the  improved  spectral capability of \xmm\ and \suzaku\, asymmetric and smeared broad Fe K lines have been observed in about 10 NS low-mass X-ray binaries \citep{disalvo09,dai09,iaria09,pandel08,bhattacharyya07,cackett08,cackett10} and were interpreted as relativistic reflection features from the inner part of the accretion disk. As shown in the self-consistent approach of \citet{dai10}, the  broad features other than that of iron in the spectrum  also support the "relativistic origin" scenario.  However, the robustness of these results with respect to pile-up effects has been questioned. \citet{ng10}, based on archival XMM observations, studied the Fe K emission of 16 NS systems, and concluded that the line profiles show no evidence of asymmetry when spectra are extracted from pile-up free regions.  
Conversely, \citet{miller10} conducted simulations for CCD based detectors to assess the impact of photon pile-up on relativistic disk line and continuum spectra, concluding that relativistic disk spectroscopy is generally robust against pile-up when the effect is modest.

Although the ``relativistic origin" is a widely accepted interpretation, alternative physical mechanisms to explain the line broadening are still debated. A broad iron line can originate from thermal Comptonization of line photons in the so-called accretion disk corona (ADC) \citep{brandt94}, or can be caused by non-thermal Comptonization in strong outflow winds \citep{laurent07}.  The predicted line profiles differ significantly from the relativistic disk line profiles.

In this \emph{Letter}, we report detection of a broad and asymmetric iron line in a spectrum of the neutron star LMXB \gx\ with high statistics and excellent energy resolution obtained with the EPIC-pn camera on board \xmm.  We discuss in detail instrumental pile-up effects and show that the relativistic profile of the line is robust against biases from photon pile-up.

\section{Observation and data analysis}

\gx\ is a typical bright ($F_{2-10~keV}\sim 150-400$~mCrab) atoll source with peculiar bursting behavior. It exhibits type-I bursts \citep{makishima83,kuulkers00} and superbursts, thought to arise from carbon shell flashes in the layers below the surface \citep{kuulkers02}. The spectrum of the source observed with the \sax\ narrow field instruments has been modeled by \citet{oosterbroek01}  with a disk-blackbody, a Comptonized component, and absorption by interstellar medium. Interestingly, a broad iron line was clearly detected but only marginally discussed by the authors. 

\gx\ was observed on September 1, 2010 by \xmm\, and simultaneously with \xte\ for about 56~ks.

The XMM-Newton Observatory \citep{jansen01} includes three
1500~cm$^2$ X-ray telescopes each with a European Photon Imaging Camera (EPIC, 0.1--15~keV) at the focus. Two of the EPIC imaging spectrometers use MOS CCDs \citep{turner01} and one uses pn CCDs \citep{strueder01}. 
Data products were reduced using the science analysis software (SAS) version 11.0. Owing to the expected high count rate from the source, the telemetry of EPIC MOS cameras was allocated to the EPIC pn camera, therefore no data from the MOS cameras are available.  We did not analyze the RGS data since they do not cover the Fe~K energy band.

The EPIC pn was operated in timing mode for the entire observation. In this mode 
only one CCD chip is operated and the data are collapsed into a one-dimensional row (4\farcm4) and read out at high speed, the second dimension being replaced by timing information. This allows a time resolution of 30~$\mu$s. We used the SAS task {\tt epfast} on the event files to correct for the effect of charge transfer inefficiency (CTI), which has been seen in the EPIC pn timing mode at high count rates\footnote[1]{http:$\slash\slash$xmm.esac.esa.int$\slash$external$\slash$xmm$\_$calibration}.
Ancillary response files were generated using the SAS task {\tt arfgen} following the recommendations in the {\it XMM-Newton SAS User guide~\footnote[2]{http:$\slash\slash$xmm.esac.esa.int}} for piled-up observations in timing mode whenever applicable. Response matrices were generated using the SAS task {\tt rmfgen}. 

Since the average EPIC pn count rate was close to the 600~\countsec\ level at which pile-up effects start to be important, we investigated in detail if there was any  pile-up before extracting the spectra. We used the SAS task {\tt epatplot}, which uses the relative ratios of single- and double-pixel events as a diagnostic tool.  These ratios deviate from standard values when there is significant pile-up in the pn camera timing mode.

The source coordinates fell in the center of column 37 of the CCD. Source events were first extracted from a 53.3\arcsec\ (13 colums) wide box centered on the source position; we call this Region~1. We also defined regions excluding the central columns. Region~2 has the central column (column 37) removed.  Region~3 has the three central columns removed (columns 36, 37, and 38).  
The relative ratios of single- and double-pixel events exhibits an energy-dependent deviation between the data and the model at $E\approxgt$~4 keV when Region~1, including all source events, is used.  
The effects of pile-up are reduced for Region~2, with the central column removed.  After removal of the three central columns, Region~3, any deviation in the ratios disappears, showing that pile-up has no significant effect.
Spectra were extracted for each of the three regions, selecting only events with PATTERN $\leq$ 4 (singles and doubles) and FLAG=0.  The energy channels were grouped by a factor of four to avoid oversampling the instrumental energy resolution. 

\section{Results}
\label{sec:results}

To model the spectrum, we used the spectral analysis package {\it XSPEC} v.12.7.0. To avoid the systematic effects discussed in \citet{dai10} that were recently addressed in the EPIC-pn calibration note\footnote[3]{ http://xmm2.esac.esa.int/docs/documents/CAL-TN-0083.pdf} we removed the energy bins below 2.4~keV and those above 9~keV. We then fitted the continuum with a model widely used for atoll LMXBs \citep{piraino07,disalvo09,barret02}: a photoelectric absorbed two-component model (Model 1) containing a {\it blackbody} and a thermal Comptonized component described by the model {\it comptt} \citep{titarchuk94}. For the photoelectric absorption, we used the XSPEC model {\it phabs}.

\begin{figure} 
\includegraphics[scale=0.39,angle=-90]{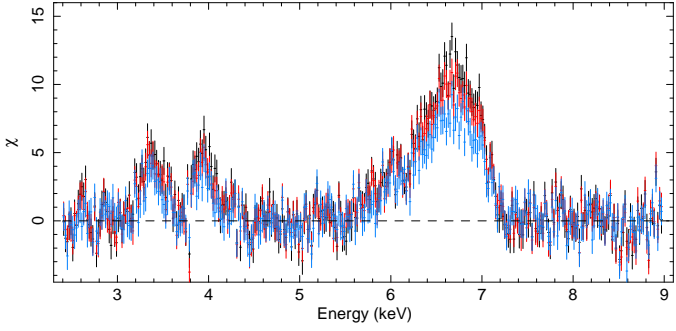}
\caption{Residuals in units of $\sigma$ in the pn energy range used with respect to the best-fitting continuum model, composed of an absorbed (phabs) diskbb and a blackbody component for the spectrum extracted in Region~1 (black), Region~2 (red) and Region~3 (blue). }    
\label{fig:residuals} 
\end{figure}

This model gave an unacceptable fit ($\chi^2/{\rm DoF} = 2621/322$) because there were several unmodeled features across the entire energy range, the most prominent being a broad profile with residuals from 5.5 to 7.3 keV and a maximum deviation of $\sim 14 \sigma$ from the best-fit continuum (see Fig.~\ref{fig:residuals}, black points).  Following the approach of \citet{disalvo09}, we added to the continuum three \emph{Gaussian} emission lines at $\sim$3.3~keV (Ar~XVIII), $\sim$3.9~keV (Ca XIX), and $\sim 6.6$~keV (Fe XXIV and/or Fe~XXV), respectively. This greatly improved the fit to $\chi^2/{\rm  DoF} = 557/313$. All these lines are significantly broader than the pn energy resolution, with \emph{Gaussian} $\sigma$ of 140~eV for Ar, 180~eV for Ca, and 290~eV for Fe. The fit was additionally  improved by adding an absorption \emph{edge} at 8.40~keV resulting in $\chi^2/{\rm DoF} =  521/311$.

Because the residuals of the Fe line profile, shown in Fig.~\ref{fig:residuals}, suggest an asymmetric shape, we replaced the Gaussian at 6.6~keV with a \emph{diskline} profile \citep{fabian89}, obtaining again a significant improvement of the fit with $\chi^2/{\rm DoF} = 451/311$.  Adding  two parameters improved the fit by $\Delta \chi^2 = 100$, corresponding to an F-test probability of chance improvement of $3 \times 10^{-15}$). 
However, the {\it comptt} parameters are poorly constrained and the high value of the optical depth, near 10, suggests that the Comptonized component is close to a Wien spectrum.  Therefore, we used a second two-component continuum model containing a \emph{diskbb} component and a simple \emph{blackbody} component (Model 2). For this model all continuum parameters are well constrained.  

\begin{figure}
\centering 
\includegraphics[scale=0.55,angle=-90]{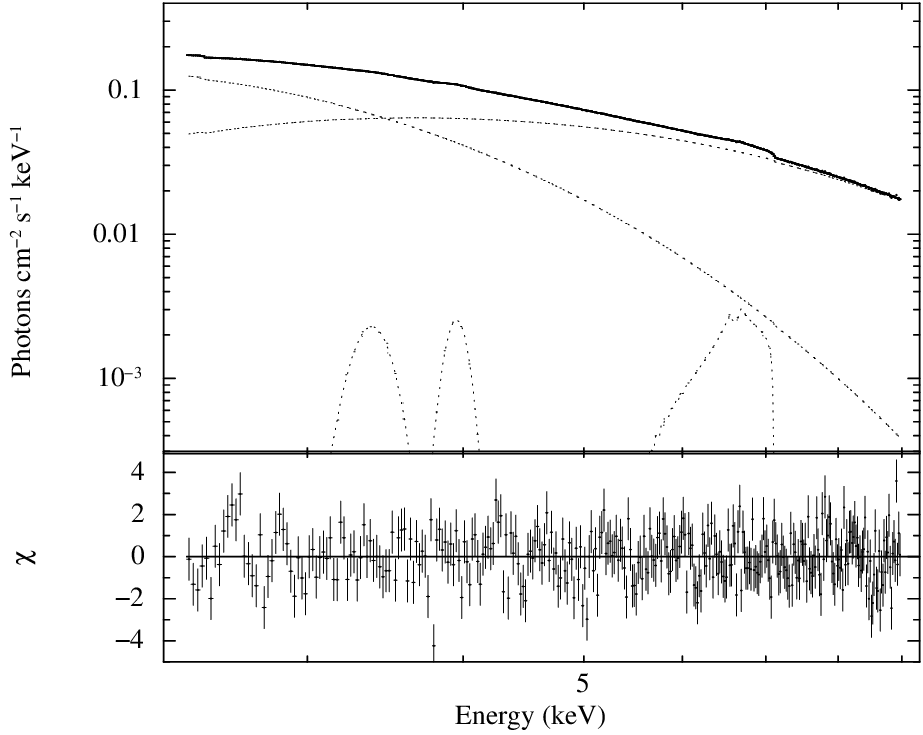}
\includegraphics[scale=0.55,angle=-90]{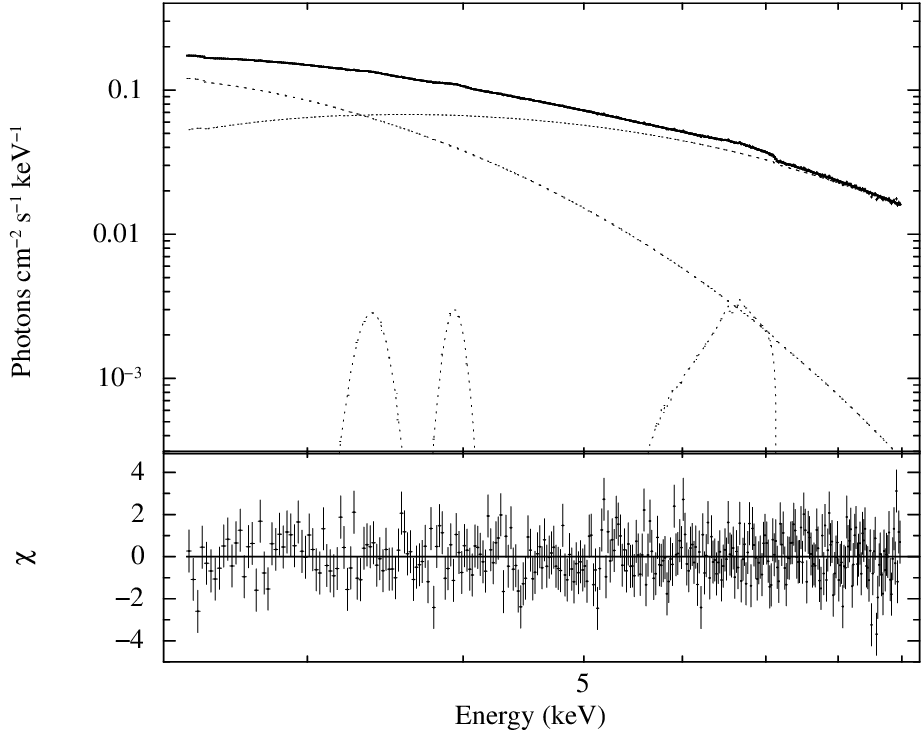}

\caption{Top: unfolded pn spectrum and individual model components for Model 2 using data from Region 1 which includes piled-up events.  Bottom: the same for Model 2, with $N_H$ fixed as described in the text, for Region 3, which excludes piled-up events. Residuals in units of $\sigma$ are shown in each panel.}
\label{fig:uf_delchi} 
\end{figure}

As for Model 1, using a \emph{diskline} profile for the iron emission line added to Model 2 significantly improves the $\chi^2/{\rm DoF}$ with respect to a \emph{Gaussian} from $549/315$ to $441/ 313$.   The $\Delta\chi^2 =108 $ for the addition of two parameters corresponds to an F-test probability of chance improvement of $1.3 \times 10^{-15}$. The \emph{diskline} parameters are well constrained, except for the outer disk radius, which we fixed after some trials at R$_{out}$=4500 ($G M / c^2$). The best-fit parameters are reported in Table~\ref{tab:specpar} and the unfolded spectrum and residuals are shown in the upper panel of Fig.~\ref{fig:uf_delchi}.

The best-fit value obtained for the energy of the iron line, $E_{line}$, suggests an ionization parameter of $\log \xi \simeq 2.5-2.7$, with $\xi$ defined as $L_X / (n r^2)$, where $L_X$ is the ionizing X-ray luminosity, $n$ the density in the reflector, and $r^2$ is the distance from the central source of emission. 

In  Column 2 of Tab.~\ref{tab:specpar}  we report the best-fit parameters of the spectrum of \gx\  for data extracted in Region 2  (without the central column 37). All spectral parameters are compatible with those obtained for data extracted from Region 1. The \emph{diskline} gives  again a better fit, with a $\Delta \chi^2 =73 $ for the addition of  two parameters, corresponding to an F-test probability of chance improvement of $2 \times 10^{-12}$. 

In the bottom  panel of the Fig.~\ref{fig:uf_delchi}, we show the unfolded pn spectrum  of \gx\  with individual components  and residuals in units of $\sigma$ for data extracted from Region 3 (without the central columns 36, 37, 38).  The best-fit parameters are reported in column 3   of Tab.~\ref{tab:specpar}. Leaving all parameters free and using the 2.4-9~keV energy range, the $N_H$ value and the \emph{diskbb} normalization slightly differ from those obtained for Regions 1 and 2, while the line parameters remain consistent. However, if we use data up to 11 keV, the $N_H$ value and the \emph{diskbb} normalization agree with those obtained for the other regions, since this spectrum is not affected by pile-up even at high energies, . Therefore, we fixed $N_H$ to the value obtained with data up to 11 keV, and again fitted the spectrum in the energy range 2.4--9~keV. All parameters are now consistent with those obtained for Region 1, as shown in Tab.~\ref{tab:specpar}, column 3. The \emph{diskline} model improves the fit, with a $\Delta \chi^2 =36$ for the addition of  two parameters and an F-test chance improvement probability of $2 \times 10^{-7}$.

As shown in Fig.~\ref{fig:residuals}, where we show the residuals in units of $\sigma$ with respect to the best-fitting continuum model, for the spectrum extracted in Region~1 (black), Region~2 (red) and Region~3 (blue), the line shape does not strongly depend on the photon pile-up fraction. To check if the low-energy lines are affected by Doppler and relativistic distortion, we replaced the Gaussian lines in Model 2 with disk lines, with all the parameters fixed at those of the iron diskline except the line centroid energy and the normalization (Model 3). We obtained an equally good fit with the elimination of two parameters, $chi^2/{\rm DoF} = 369/316 $ (see the rightmost column of Table~\ref{tab:specpar}).

\begin{table}[t]
\caption[]{The best-fit parameters for spectra of \gx\ observed with the \xmm~EPIC-pn using different models and data extraction.  The continuum used in both models reported here, Models 2 and 3, consists of the sum of a blackbody and a multicolor disk blackbody. For Model 2 we report results obtained using data extracted from three different regions. In Model 3, the low-energy lines are modeled with diskline profiles, with all parameters fixed at those of the iron diskline except the line centroid energy and the normalization. The blackbody luminosity is given in units of $L_{37} / D_{10}^2$, where $L_{37}$ is the bolometric luminosity in units of $10^{37}$~ergs/s and $D_{10}$ the distance to the source in units of 10~kpc. 
The diskbb normalization is given in units of $(R_{in}/ D_{10})^2  cos\theta $ where $R_{in}$ is an apparent disk radius in km  and $\theta$ the angle of the disk. The line intensity $I_{line}$ is given in units ph cm$^{-2}$ s$^{-1}$. $\Delta \chi$  is referred to the switch from a Gaussian  to a \emph{diskline }to model the iron line,  and the Prob$_{F-test}$ the relative probability of chance improvement given by the F-test. Flux is calculated in the $2-10$ keV band, and  given in units erg cm$^{-2}$ s$^{-1}$. Uncertainties are given at  a $90\%$ confidence level.} 
\label{tab:specpar}
\begin{center}
\begin{tabular}{l|lll|l}
\scriptsize\textbf{Parameter}  						& \multicolumn{3}{c}{\scriptsize\textbf{Model 2}}											&\scriptsize\textbf{Model 3} \\
\hline
											&\scriptsize\textbf{Region 1} 		& \scriptsize\textbf{Region 2} 		&\scriptsize\textbf{Region 3}	& \scriptsize\textbf{Region 3}   \\
 \hline
\scriptsize$N_{\rm H} \rm [10^{22} cm^{-2}$] 			& \scriptsize$1.7 \pm 0.1$		&\scriptsize $1.7 \pm 0.1$ 		 &\scriptsize$1.8 fixed$ 		&\scriptsize$1.8 fixed$ \\
\scriptsize$kT_{\rm bb} \rm [keV]$          				&\scriptsize $1.70 \pm 0.02$ 		&\scriptsize $1.68 \pm 0.02$		 &\scriptsize$1.62 \pm 0.01$ 	&\scriptsize$1.60 \pm 0.01$\\
\scriptsize$L_{\rm bb} $						        &\scriptsize $4.4 \pm 0.1$		&\scriptsize $4.4 \pm 0.1$ 		 &\scriptsize$4.4 \pm 0.1$	&\scriptsize$4.45 \pm 0.05$\\
\scriptsize$kT_{\rm dbb} \rm [keV]$ 					&\scriptsize $0.93 \pm 0.03$    	&\scriptsize $0.92 \pm 0.04$		 &\scriptsize$0.89 \pm 0.02$	&\scriptsize$0.85 \pm 0.02$\\
\scriptsize$N_{\rm dbb}$							&\scriptsize $228\pm37$  		&\scriptsize $234_{-41}^{+50}$ 		&\scriptsize$278_{-22}^{+23}$	&\scriptsize$321_{-25}^{+24}$\\
\scriptsize$E_{line}$ (keV) 						&\scriptsize $3.37\pm  0.02$		&\scriptsize $3.38\pm  0.02$		&\scriptsize$3.38\pm0.03$	&\scriptsize$3.32\pm0.03$   \\
\scriptsize$\sigma \rm [keV]$						&\scriptsize $0.12 \pm 0.03$		&\scriptsize $0.11 \pm 0.03$ 		 &\scriptsize$0.09 \pm 0.04$	&\scriptsize \\
\scriptsize$I_{line}$ ($10^{-4}$)						&\scriptsize $9\pm 2$ 			&\scriptsize $8\pm 2$ 			 &\scriptsize$8\pm 2$		&\scriptsize$13\pm 3$\\
\scriptsize EqW (eV)      							&\scriptsize $5 \pm1$    			&\scriptsize $5 \pm1 $ 			 &\scriptsize$5 \pm1$		&\scriptsize$8 \pm2 $ \\
\scriptsize$E_{line}$ (keV) 						&\scriptsize $3.95 \pm 0.02$ 		&\scriptsize $3.95 \pm 0.02$ 		 &\scriptsize $3.94 \pm 0.02$	&\scriptsize$3.91 \pm 0.03$ \\
\scriptsize$\sigma \rm [keV]$ 						&\scriptsize $0.08\pm 0.03$    		&\scriptsize $0.06\pm 0.03$ 		 &\scriptsize$0.07\pm 0.04$	&\scriptsize\\
\scriptsize$I_{line}$ ($10^{-4}$) 						&\scriptsize $6 \pm 1$  			&\scriptsize $5 \pm 1$ 			 &\scriptsize $6 \pm 2$ 		&\scriptsize$12 \pm 2$ \\
\scriptsize EqW (eV)           						&\scriptsize $ 5\pm3 $			&\scriptsize $ 4\pm3 $			 &\scriptsize$ 5\pm3$		&\scriptsize$ 10\pm2 $  \\  
\scriptsize$E_{line}$ (keV) 						&\scriptsize $6.62\pm 0.02$ 		&\scriptsize $6.60\pm 0.02$ 		 &\scriptsize$6.60\pm 0.03$	&\scriptsize$6.61\pm 0.02$ \\
\scriptsize$I_{line}$ ($10^{-3}$) 						&\scriptsize $2.4 \pm 0.2$ 		&\scriptsize $2.6 \pm 0.2$ 		 &\scriptsize$2.9_{-0.2}^{+0.5}$&\scriptsize$2.6 \pm 0.2$\\
\scriptsize$EW \rm [eV]$          					&\scriptsize $57\pm5 $            		&\scriptsize $63\pm5 $			 &\scriptsize$ 70\pm6$		&\scriptsize$ 64\pm5 $\\
\scriptsize$Betor$        							&\scriptsize $-2.61 \pm 0.03$  		&\scriptsize $-2.62 \pm 0.07$ 		 &\scriptsize$-2.6 \pm 0.1$ 	&\scriptsize$-2.54 \pm 0.09$     \\  \scriptsize$R_{in}$ ($G M / c^2$)  					&\scriptsize $ 25\pm4$ 			&\scriptsize $ 25_{-9}^{+5} $		 &\scriptsize$ 28_{-12}^{+9} $ 	&\scriptsize$ 27_{-14}^{+7} $  \\  \scriptsize Incl (deg)      							&\scriptsize $37 \pm 2$    	  	&\scriptsize $39 \pm 2$  			 &\scriptsize$40_{-3}^{+4}$ 	&\scriptsize$39_{-3}^{+4}$ \\  
\scriptsize$\Delta\chi^2$ 							&\scriptsize $108$	   		  	&\scriptsize $73$ 				 &\scriptsize$36$ 			&\scriptsize \\
\scriptsize$Prob_{F-test}$						 	&\scriptsize $1.3 \times10^{-15}$	&\scriptsize $2 \times 10^{-12}$	 &\scriptsize$2 \times 10^{-7}$	&\scriptsize\\
\scriptsize Flux ($10^{-9}$ )  				 		&\scriptsize $4.53 \pm0.05 $		&\scriptsize $4.49 \pm 0.05 $		 &\scriptsize$4.48 \pm0.06 $	&\scriptsize$4.48 \pm0.06 $\\
\hline
\scriptsize$\chi^2$ (dof)							&\scriptsize $441~(313)$			&\scriptsize $389~(313)$			 &\scriptsize$352~(314)$		&\scriptsize$369~(316)$\\
\hline
\noalign{\smallskip}
\end{tabular}
\end{center}
\end{table}
\normalsize

\section{Discussion}
\label{sec:discussion}

We reported the detection of broad emission lines at $\sim$3.3~keV, $\sim$3.9~keV and $\sim$6.6~keV, corresponding to Ar XVIII, Ca XIX, and most likely a blend of Fe XXIV and Fe XXV, in high-quality {\xmm} X-ray spectra of the LMXB and atoll source {\gx} obtained when the source was in a relatively low-intensity state. A broad, $\sigma\sim$1.0 keV, iron line feature was previously reported by \citet{oosterbroek01}, who simply mentioned that the line might originate from reflection from an accretion disk or from Compton broadening in the external parts of a $\sim$3 keV corona.  We showed that the disk-reflection scenario is consistent with the high-quality data from the {\xmm} EPIC-PN. The iron line's broadening and shape are therefore affected by Doppler and relativistic effects at the inner edge of the accretion disk. In addition, Compton broadening is expected due to the hot ionized disk \citep{ross07}. Such a component does not affect  the skewness of the line, however, and indeed we have verified that even  including  Compton broadening, relativistic smearing is required with high statistical significance.

In our study of the spectral properties, we carefully investigated effects caused by instrumental photon pile-up.  We conclude that the line shape does not significantly depend on the photon pile-up fraction.  Our results agree with the conclusion of Miller et al.\ (2010) that relativistic disk spectroscopy is generally robust against pile-up when this effect is modest.
We measured an iron line rest-frame energy, $E\sim 6.61$ keV that is compatible with line emission from Fe XXIV and Fe XXV, showing that the reflecting matter is ionized and leading to an estimate for the ionization parameter of $\log \xi \simeq 2.4-2.7$.  The line appears to be produced at radii extending inward to $R_{in} = (25 \pm 4) R_g$,  
where $R_g = G M/c^2$ is the gravitational radius.  Because $R_g \sim 1.5$~km for a $1 M_\odot$ neutron star, the inner radius is then $R_{in} = (50 \pm 8) m_{1.4}$~km, where $m_{1.4}$ is the neutron star mass in units of $1.4\;M_\odot$.  For reasonable values of the neutron star mass, our results indicates that the disk is truncated quite far from the NS surface. The apparent inner radius of the soft disk blackbody emission gives a radius of $8\pm2 $~km, calculated for a distance to the source of 4.5 kpc \citep{kuulkers00}. Note that this value is thought to underestimate the true inner radius by a factor of more than 2 \citep{merloni00}.  The power law dependence of the disk emissivity, $r^{betor}$, is parametrized by the index $betor$, which is $-(2.61 \pm 0.03)$, indicating that the disk emission is dominated by irradiation by a central source (see \citealt{fabian89}). Finally, the disk inclination with respect to the line of sight is also well determined, $i = (37 \pm 2)$ degrees. This agrees well with the intermediate inclinations, between 20 and 65 degrees, found in other sources, as indeed expected from selection criteria.
 
Signatures of a disk reflection are also the discrete features present at soft X-rays from Ar XVIII, and Ca XIX, respectively.  We note that the scenario emerging from this work on \gx\ is coherent with results obtained for other sources and is in particular very similar to the case of 4U 1705-44 \citep{disalvo09,dai10}.

We focused on the robust detection of the relativistic iron line in the source, taking into account systematic instrumental effects caused by pile-up. Additional studies based on more physical self-consistent reflection models are in progress. We also aim to present more extensive spectral-timing studies based on the simultaneous RXTE observation and on the XMM RGS in the near future.

\begin{acknowledgements}

We thank the XMM and RXTE teams for the rapid scheduling of the  observations of \gx.
The T\"ubingen group was supported by DLR (grant DLR 50~OR~0702) and EE by Initial Training Network ITN 215212: Black Hole Universe funded by the European Community. 

\end{acknowledgements}

\bibliographystyle{aa}

\bibliography{ref}

\end{document}